\documentclass[preprintnumbers, floatfix,letterpaper,aps,prd,epsfig,nofootinbib,
twocolumn
]{revtex4-1}
\usepackage{bm,graphicx,dcolumn,epstopdf,epsf, latexsym,mathbbol, amssymb,amsmath,color,slashed, mathrsfs,mathcomp,simplewick}
\usepackage[center]{subfigure}
\usepackage{multirow}
\usepackage{makecell}
\usepackage[colorlinks,linkcolor=blue,citecolor=blue,urlcolor=blue]{hyperref}

\begin{document}
\allowdisplaybreaks
 \newcommand{\bq}{\begin{equation}}
 \newcommand{\eq}{\end{equation}}
 \newcommand{\bqn}{\begin{eqnarray}}
 \newcommand{\eqn}{\end{eqnarray}}
 \newcommand{\nb}{\nonumber}
 \newcommand{\lb}{\label}
 \newcommand{\f}{\frac}
 \newcommand{\p}{\partial}
\newcommand{\PRL}{Phys. Rev. Lett.}
\newcommand{\PLB}{Phys. Lett. B}
\newcommand{\PRD}{Phys. Rev. D}
\newcommand{\CQG}{Class. Quantum Grav.}
\newcommand{\JCAP}{J. Cosmol. Astropart. Phys.}
\newcommand{\JHEP}{J. High. Energy. Phys.}
\newcommand{\red}{\textcolor{black}}
%

\title{Gravitational wave constraints on Lorentz and parity violations in gravity: High-order spatial derivative cases}

\author{Cheng Gong ${}^{a, b, c}$}

\author{Tao Zhu${}^{b, c}$}
\email{corresponding author: zhut05@zjut.edu.cn}

\author{Rui Niu${}^{d, e}$}

\author{Qiang Wu${}^{b, c}$}

\author{Jing-Lei Cui${}^{a}$}

\author{Xin Zhang${}^{a}$}
\email{zhangxin@mail.neu.edu.cn}

\author{Wen Zhao${}^{d, e}$}
\email{wzhao7@ustc.edu.cn}

\author{Anzhong Wang${}^{f}$}
 \email{anzhong$\_$wang@baylor.edu}
\affiliation{
${}^{a}$ Department of Physics, College of Sciences, Northeastern University, Shenyang 110819, China\\
${}^{b}$ Institute for Theoretical Physics and Cosmology, Zhejiang University of Technology, Hangzhou, 310032, China\\
${}^{c}$ United Center for Gravitational Wave Physics (UCGWP), Zhejiang University of Technology, Hangzhou, 310032, China\\
${}^{d}$ CAS Key Laboratory for Research in Galaxies and Cosmology, Department of Astronomy, University of Science and Technology of China, Hefei 230026, China; \\
${}^{e}$ School of Astronomy and Space Sciences, University of Science and Technology of China, Hefei, 230026, China;\\
${}^{f}$ GCAP-CASPER, Physics Department, Baylor University, Waco, Texas 76798-7316, USA}

\date{\today}

\begin{abstract}

High-order spatial derivatives are of crucial importance for constructing the low energy effective action of a Lorentz or parity violating theory of quantum gravity. One example is the Ho\v{r}ava-Lifshitz gravity, in which one has to consider at least the sixth-order spatial derivatives in the gravitational action, in order to make the theory power-counting renormalizable. In this paper, we consider the Lorentz and parity violating effects on the propagation of GWs due to the fifth and sixth-order spatial derivatives respectively. For this purpose we calculate the corresponding Lorentz and parity violating waveforms of GWs produced by the coalescence of compact binaries. By using these modified waveforms, we perform the full Bayesian inference with the help of the open source software \texttt{BILBY} on the selected GW events of binary black hole (BBH) and binary neutron stars (BNS) merges in the LIGO-Virgo catalogs GWTC-1 and GWTC-2. Overall we do not find any significant evidence of Lorentz and parity violation due to the fifth and sixth-order spatial derivatives and thus place lower bounds on the energy scales $M_{\rm LV} > 2.4 \times 10^{-16} \; {\rm GeV}$ for Lorentz violation and $M_{\rm PV} > 1.0 \times 10^{-14} \; {\rm GeV}$ for parity violation at 90\% confidence level. Both constraints represent the first constraints on the fifth- and sixth-order spatial derivative terms respectively in the framework of spatial covariant gravity by using the observational data of GWs.

\end{abstract}


\maketitle
\section{Introduction}
\renewcommand{\theequation}{1.\arabic{equation}} \setcounter{equation}{0}

The detection of gravitational waves (GW) from individual compact binary sources(CBC) by the LIGO/Virgo collaboration over the past few years has started a new era in gravitational physics \cite{gw150914,gw-other, gw170817, LIGOScientific:2017ycc, gwtc1, gwtc2, LIGOScientific:2021djp}. These signals are consistent with GWs produced by the coalescence of compact binaries, as predicted by Einstein's general relativity (GR). These GWs also provide strong evidences  in the strong field and highly dynamical regime in which GR's predictions are well tested.
Up to now, GW have become an important tool in testing GR, and it is a very important topic in the era of GW astronomy \cite{gw150914-testGR, gw170817-testGR, gw170817-speed, testGR_GWTC1, testGR_GWTC2}. 

Although GR is the most successful gravity theory at present, there are difficulties in explaining some problems, including(e.g., singularity, quantization, etc), and observationally (e.g., dark matter, dark energy, etc).  Therefore, in order to solve the above problems, various modified theories of gravity have been proposed \cite{MG1, MG2, MG3, MG4}. Therefore, the tests of the modified gravities are essential to confirm the final theory of gravity. 

On the other hand, the Lorentz invariance is one of the fundamental principle of GR. However, when one considers the quantization of the gravity, such invariance could be violated at high energy regime. In this sense, the Lorentz symmetry can be treated as a approximate symmetry, which emerges at low energies and is violated at higher energies. With these thoughts, a lot of modified gravity theories have been proposed, such as the Ho\v{r}ava-Lifshitz theories of quantum gravity (see Refs. \cite{horava1, horava2, horava3, horava4, horava5, horava6} for examples), Einstein-\AE{}ther theory \cite{davi, d1, jacobson_einstein-aether_2008, li, bat}, spatial covariant gravities \cite{Gao:2020qxy, Gao:2020yzr, Gao:2019twq, Gao:2019liu, Joshi:2021azw}, and the standard model extension \cite{SME1, SME2, SME3, SME4, SME5, SME6}. When the Lorentz symmetry is violated, the associated gauge symmetry of gravity can be broken from the general covariance down to the three dimensional spatial covariance. Then the gravitational action can be constructed from those terms which respect the spatial diffeomorphisms. In such framework, high-order spatial derivatives are of crucial importance for constructing the low energy effective action of a Lorentz or parity violating theory of quantum gravity. For example, in Ho\v{r}ava-Lifshitz gravity \cite{horava1, horava6}, one has to consider at least the sixth-order spatial derivatives in the gravitational action, in order to make the theory power-counting renormalizable. 

Breaking the Lorentz symmetry in gravitational interaction also provides a natural way to incorporate the parity violation terms into the theory. With spatial covariance, the parity violation can be achieved by including the odd-order spatial derivatives into the gravitational action \cite{horava5, Gao:2019liu}. Several parity violating scalar-tensor theories can also be mapped to the spatial covariant framework by imposing the unitary gauge \cite{Gao:2019liu, Gao:2020yzr, cs1, chiral_ST}. In these theories, the parity violating terms can also be written in the form of high-order odd-order spatial derivatives. Various implications of parity violations on the gravitational waves have been extensively explored in  \cite{cs_review, cs3, cs4, cs5, chiral_ST1, PGW1, PGW2, PGW3, Fu:2020tlw,add1, add2, add3}. Note that in \cite{Conroy, Li:2020xjt,Li:2021wij, Li:2021mdp},  parity-violating effects induced by lower dimension operators have also been proposed and their phenomenological effects have been explored.

Both the Lorentz and parity violations in gravity can introduce deviations from GR in the propagation of GWs. When the Lorentz symmetry is violated, the conventional linear dispersion relation of GWs can be modified into a nonlinear one, which in turn changes the phase velocities of GWs at different frequencies \cite{Mirshekari:2011yq}. For parity violation, in general it causes the polarization propagation velocity and amplitude damping of the left and right hands of GW to be different, which leads to the velocity and amplitude birefringence, respectively \cite{waveform}. 

Recently, a lot of tests on both the Lorentz symmetry and parity symmetry of gravitational interaction have been carried out by using the observational data from GW events in LIGO-Virgo catalogs \cite{Okounkova:2021xjv, Hu:2020rub, CS_gb, sai_wang, tanaka, yi-fan1, Wang:2021gqm}. In all these works, only the effects of Lorentz and parity violation due to the leading spatial derivatives have been considered. In this paper, we study in detail the GW effects of the Lorentz violation due to the sixth-order derivatives and parity violation due to fifth-order derivatives in the gravitational action. Decomposing the GWs into the left-hand and right-hand circular polarization modes, we find that the GW effects of both the Lorentz and parity violations can be explicitly presented by the modifications in the GW amplitude and phase. We also mapped such amplitude and phase modification to the parametrized description of Lorentz and parity violating waveforms proposed in \cite{waveform}.  With the modified waveform, we perform the full Bayesian inference with the help of the open source software {\bf BILBY} on the selected GW events in the LIGO-Virgo catalogs GWTC-1 and GWTC-2. From our analysis, we do not find any signatures of Lorentz violation due to the sixth-order derivatives and parity violation due to the fifth-order spatial derivatives in the gravitational action in the GW data except GW190521. We thus place lower bounds on the energy scale $M_{\rm LV} > 2.4 \times 10^{-16} \; {\rm GeV}$ for Lorentz violation and $M_{\rm PV} > 1.0 \times 10^{-14} \; {\rm GeV}$ for parity violation at 90\% confidence level. Both constraints represent the first constraints on the fifth- and sixth-order spatial derivative terms respectively in the spatial covariant framework by using the observational data of GWs.

This paper is organized as follows. In the next section, we present a brief introduction of GWs and waveform in Lorentz and parity violating gravities and then discuss Lorentz violating dispersion relations and parity violating birefringences. We also calculate in details the waveforms of GWs produced by the coalescence of compact binary systems and particularly focus on the deviations from those in GR. In Sec. III, we present two examples of modified gravity which naturally involved high-order spatial derivatives in the spatial covariant framework. In Sec.~IV, we present the basic statistical framework of Bayesian analysis used in this work and report the results of constraints on the Lorentz and parity violation from the Bayesian analysis in Sec. V. We finish with concluding remarks and discussion in Sec.~VI.

Throughout this paper, the metric convention is chosen as $(-,+,+,+)$, and greek indices $(\mu,\nu,\cdot\cdot\cdot)$ run over $0,1,2,3$ and latin indices $(i, \; j,\;k)$ run over $1, 2, 3$. We choose the units to $G =c=1$.

\section{Waveform of GWs with Lorentz and parity violating effects}
\renewcommand{\theequation}{2.\arabic{equation}} \setcounter{equation}{0}

In this section, we present a brief review the waveform of GWs with Lorentz and parity violations. Both the Lorentz and parity violations can modify the damping rates or phase velocities of the two tensorial modes of GWs. For our purpose we restrict our attention to the cases with Lorentz and parity violations due to high-spatial derivative terms in gravitational actions and adopt the general parametrized framework developed in \cite{waveform} for describing the effects of Lorentz and parity violating effects in the propagation of GWs. 

\subsection{GWs in parity and Lorentz violating gravities}

We consider the GWs propagating on a homogeneous and isotropic background. The spatial metric in the flat Friedmann-Robertson-Walker universe is written as
\bqn
g_{ij} = a(\tau) (\delta_{ij} + h_{ij}(\tau, x^i)), 
\eqn
where $\tau$ denotes the conformal time, which relates to the cosmic time $t$ by $dt =a d\tau$, and $a$ is the scale factor of the universe. Throughout this paper, we set the present scale factor $a_0 =1$. $h_{ij}$ denotes the GWs, which represents the transverse and traceless metric perturbations, i.e, 
\bqn
\partial^i h_{ij} =0 = h^i_i.
\eqn
To study the Lorentz and parity violating effects on GWs, it is convenient to decompose the GWs into the circular polarization modes. To study the evolution of $h_{ij}$, we expand it over spatial Fourier harmonics,
 \bqn
 h_{ij}(\tau, x^i) = \sum_{A={\rm R, L}} \int \frac{d^3k}{(2\pi)^3} h_A(\tau, k^i) e^{i k_i x^i} e_{ij}^A(k^i),\nb\\
 \eqn
 where $e_{ij}^A$ denote the circular polarization tensors and satisfy the relation
 \bqn
 \epsilon^{ijk} n_i e_{kl}^A = i \rho_A e^{j A}_l,
 \eqn
 with $\rho_{\rm R} =1$ and $\rho_{\rm L} = -1$. We find that the propagation equations of these two modes are decoupled, which can be casted into the parametrized form \cite{waveform}
 \bqn\lb{eom_A}
 h''_A + (2+\bar \nu + \nu_A) \mathcal{H} h'_A + (1+\bar \mu+ \mu_A) k^2 h_A=0,\nb\\
 \eqn
 where a prime denotes the derivative with respect to the conformal time $\tau$ and $\mathcal{H} =a'/a$. 
 
 In the above parametrization, the new effects arising from theories beyond GR are fully characterized by four parameters: $\bar \nu$, $\bar \mu$, $\nu_A$ and $\mu_A$. The parameters $\nu_A$ and $\mu_A$ label the effects of the parity violation, and $\bar \nu$ and $\bar \mu$ describe the effects of other possible modifications which are not relevant to parity violation. In these four parameters, $\mu_A$ and $\bar \mu$ determine the speed of GWs, while $\nu_A$ and $\bar \nu$ determine the damping rate of GWs during their propagation.  As shown in  \cite{waveform}, such parametrization provides an unifying description for the low-energy effective description of GWs in generic modified gravities, including Chern-Simons modified gravity \cite{cs_review}, ghost-free parity violating scalar-tensor theory \cite{chiral_ST, chiral_ST1, Gao:2019liu}, symmetric teleparallel equivalence of general relativity \cite{Conroy}, Ho\v{r}ava-Lifshitz gravities \cite{horava1, horava2, horava3, horava4, horava5, horava6}, the Nieh-Yan modified teleparallel gravity \cite{Li:2020xjt, Li:2021wij}.

With regard to the parity violation, the parameter $\mu_A$ leads to different velocities of left- and right-hand circular polarizations of GWs, so the arrival times of the two circular polarization modes could be different, which is the phenomenon of velocity birefringence. The parameter $\nu_A$, on the other hand, leads to different damping rates of  left- and right-hand circular polarizations of GWs, so  the amplitude of left-hand circular polarization of GWs will increase (or decrease) during the propagation, while the amplitude for the right-hand modes will decrease (or increase). This is the  phenomenon of amplitude birefringence. It is also worth noting that for  large numbers of parity violating gravities, both  parameters $\nu_A$ and $\mu_A$ are frequency dependent. 

For the effects which are not relevant to parity violation, the parameters $\bar \nu$ and $\bar \mu$ can arise either from a modified gravity with general covariance or a modified theories in the framework of spatial covariant gravities. For the former case, both $\bar \nu$ and $\bar \mu$ can be frequency-independent, as shown in Table.~I of \cite{waveform} for a few of modified gravities. For the later case, the theories only respect the spatial covariance, not the usual general covariance of four dimensional spacetime. In this case, there exists a preferred timelike direction which violates the temporal diffeomorphism of the theories, thus breaks the Lorentz symmetry of the theories as well. In these kind of theories, the parameters $\bar \nu$ and $\bar \mu$ can be frequency dependent. In this paper, in order to study the Lorentz violating effects on GWs,  we only focus on the later case.

As analyzed in the above, here we can assume all the four parameters are frequency dependent, the two parameters $\nu_A$ and $\mu_A$ are effects arising from parity violations and the other two parameters $\bar \nu$ and $\bar \mu$ are effects from Lorentz violation. In this sense, we parametrize these four parameters in the following forms \cite{waveform}
\bqn
\mathcal{H} \bar{\nu} &=&\left[\alpha_{\bar{\nu}}(\tau)\left(k / a M_{\mathrm{LV}}\right)^{\beta_{\bar{\nu}}}\right]', \\
 \bar{\mu}&=&\alpha_{\bar{\mu}}(\tau)\left(k / a M_{\mathrm{LV}}\right)^{\beta_{\bar{\mu}}}, \\
\mathcal{H} \nu_{\mathrm{A}} &=&\left[\rho_{\mathrm{A}} \alpha_{\nu}(\tau)\left(k / a M_{\mathrm{PV}}\right)^{\beta_{\nu}}\right]^{\prime}, \\
\mu_{\mathrm{A}}&=&\rho_{\mathrm{A}} \alpha_{\mu}(\tau)\left(k / a M_{\mathrm{PV}}\right)^{\beta_{\mu}},
\eqn
where $\beta_{\bar \nu}$ and $\beta_{\bar \mu}$ are arbitrary even numbers and $\beta_\nu$ and $\beta_\mu$ are arbitrary odd numbers. $\alpha_{\bar \nu}$, $\alpha_{\bar \mu}$, $\alpha_\nu$, and $\alpha_\mu$ are  arbitrary functions of time. These parameters can only be determined given a specific model of modified gravities. For the GW events at local Universe, these two functions can be approximately treated as constant, i.e. ignoring their time-dependence.

 \subsection{Waveforms of GWs with parity and Lorentz violations}
 
 With the above parametrization of Lorentz and parity violating effects on the propagation of GWs, one can derive their explicit GW waveform by solving the equation of motion (\ref{eom_A}). It is shown in \cite{waveform} that the amplitude and phase modifications to the GR-based waveform due to both Lorentz and parity violating effects can be written in the following parametrized form
 \bqn
 \tilde h_A(f) = \tilde h_A^{\rm GR} (1+ \rho_A \delta h_1 + \delta h_2) e^{i (\rho_A \delta \Psi_1 + \delta \Psi_2)},\nb\\
 \label{2.10}
 \eqn
 where $ \tilde h_A^{\rm GR} $ is the corresponding waveform in GR, and its explicit form can be found in the previous works \cite{waveform}.  The amplitude corrections $\delta h_1$ and $\delta h_2$ are caused by the parameters $\nu_A$ and $\bar \nu$ respectively, while the  phase corrections $\delta \Psi_1$ and $\delta \Psi_2$ are caused by the parameters $\mu_A$ and $\bar \mu$ respectively. The explicit forms of $\delta h_1$, $\delta h_2$, $\delta \Psi_1$, and $\delta \Psi_2$ are given respectively by
 \bqn
 \delta h_1 = - \frac{1}{2} \left(\frac{2 \pi f}{M_{\rm PV}}\right)^{\beta_\nu}\Big[\alpha_\nu(\tau_0) - \alpha_\nu(\tau_e) (1+z)^{\beta_\nu}\Big], \nb\\
 \;\; \\
 \delta h_2 = - \frac{1}{2} \left(\frac{2 \pi f}{M_{\rm LV}}\right)^{\beta_{\bar \nu}}\Big[\alpha_{\bar \nu}(\tau_0) - \alpha_{\bar \nu}(\tau_e) (1+z)^{\beta_{\bar \nu}}\Big], \nb\\
 \;\; \\
 \delta \Psi_1 = 
 \begin{cases}\label{formula 2.13}
 \frac{(2/M_{\rm PV})^{\beta_\mu}}{\beta_\mu+1} (\pi f)^{\beta_\mu +1} \int_{t_e}^{t_0} \frac{\alpha_\mu}{a^{\beta_\mu+1}}dt, & \;\; \beta_\mu \neq -1, \\
 \frac{M_{\rm PV}}{2} \ln u \int_{t_e}^{t_0} \alpha_{\bar \mu} dt, & \;\; \beta_\mu = -1, \nb
 \end{cases}
 \;\; \\
 \eqn
 and 
 \bqn
 \delta \Psi_2 = 
 \begin{cases}\label{formula 2.14}
  \frac{(2/M_{\rm LV})^{\beta_{\bar \mu}}}{\beta_{\bar \mu}+1} (\pi f)^{\beta_{\bar \mu} +1} \int_{t_e}^{t_0} \frac{\alpha_{\bar \mu}}{a^{\beta_{\bar \mu}+1}}dt, & \;\; \beta_{\bar \mu} \neq -1, \\
 \frac{M_{\rm LV}}{2} \ln u \int_{t_e}^{t_0} \alpha_\mu dt, & \;\; \beta_{\bar \mu} = -1, \nb
 \end{cases}
 \;\;\\
 \eqn
 where $t_e$ ($t_0$) denotes the emitted (arrival) time for a given GW event, $z=1/a(t_e) -1 $ is the cosmological redshift, $f$ is the GW frequency at the detector, and $u=\pi {\cal M} f $, where  ${\cal M}$ is the measured chirp mass of the binary system. 
 
 The circular polarization modes $\tilde h_{\rm R}$ and $\tilde h_{\rm L}$ relate to the modes $\tilde h_{+}$ and $h_{\times}$ via
 \bqn
 \tilde h_{+} = \frac{\tilde h_{\rm L} + \tilde h_{\rm R}}{\sqrt{2}}, \\
  \tilde h_{\times} = \frac{\tilde h_{\rm L} - \tilde h_{\rm R}}{\sqrt{2}i}. 
 \eqn
 Thus the waveforms for the plus and cross modes become
 \bqn
&\tilde{h}_{+}=\tilde{h}_{+}^{\mathrm{GR}}\left(1+\delta h_{2}+i \delta \Psi_{2}\right)-\tilde{h}_{\times}^{\mathrm{GR}}\left(i \delta h_{1}-\delta \Psi_{1}\right), \nb\\
&& \;\; \lb{hplus}\\
&\tilde{h}_{\times}=\tilde{h}_{\times}^{\mathrm{GR}}\left(1+\delta h_{2}+i \delta \Psi_{2}\right)+\tilde{h}_{+}^{\mathrm{GR}}\left(i \delta h_{1}-\delta \Psi_{1}\right).\lb{htimes}\nb\\
 \eqn
 
 As we mentioned, the coefficients $\alpha_{\nu}$($\beta_{\nu}$), $\alpha_{\bar \nu}$($\beta_{\bar \nu}$), $\alpha_\mu$ ($\beta_{\mu}$), and $\alpha_{\bar \mu}$ ($\beta_{\bar \mu}$) depend on specific modified gravities. Different numbers of $\beta_{\nu, \bar \nu, \mu, \bar \mu}$ correspond to different effects. In this paper, we consider two distinct effects, one from Lorenz violation and the other from parity violation.
 
 \subsubsection{Lorentz violating dispersion relations}
 
 Except $\beta_{\bar \mu}=0=\beta_{\bar \nu}$ and $\beta_{\bar \mu}=-2=\beta_{\bar \nu}$, the coefficients $\alpha_{\nu}$ ($\beta_{\nu}$) and $\alpha_{\mu}$ ($\beta_{\mu}$) represent the Lorentz violating effects on the propagation of GWs. Among these parameters, $\alpha_{\nu}$ ($\beta_{\nu}$) label the amplitude corrections to the waveform of the GWs, while $\alpha_{\mu}$ ($\beta_{\mu}$) is related to the phase corrections. In general, the amplitude and phase corrections can be related to each other. Since the GW detectors are more sensitive to the phase, rather than the amplitude of GWs from all the events detected by LIGO and Virgo, in this paper let us only focus on those effects that can lead to the phase corrections to the GW waveform. Thus hereafter we can set $\alpha_{\nu}=0$. Then the Lorentz violating effects associated with parameters $\alpha_{\mu}$ ($\beta_{\mu}$) can be related to a Lorentz-violating dispersion relation,
 \bqn
 \omega_k^2 = k^2 \Big[ 1+ \alpha_{\bar \mu} \left(\frac{k}{a M_{\rm LV}}\right)^{\beta_{\bar \mu}} \Big],
 \eqn
 where $\omega_k^2$ is the energy of the GWs. The waveform of GWs with the above Lorentz-violating dispersion relation has been also derived  in \cite{Mirshekari:2011yq} and then their possible constraints with ground-based and space-based detectors has been analyzed in the framework of  Fisher matrix. With the GW events detected by LIGO-Virgo, the gravitational wave constraints on the Lorenz-violating dispersion relations for $\beta_{\bar \mu} =-2, \; -1.5, \;-1, \; -0.5, \; 0.5, \; 1, \; 1.5, \;2$ have been obtained by comparing the modified waveforms with GW data \cite{testGR_GWTC2, testGR_GWTC1}. With these constraints, one is able to infer the GW constraints on the corresponding Lorentz violating gravities. In this paper, we shall consider the case with $\beta_{\bar \mu}=4$ and derive the corresponding constraints by using the signals of GW events detected by LIGO-Virgo. As we will show later, the Lorentz-violating dispersion relation with $\beta_{\bar \mu}=4$ is related to the sixth-order spatial derivatives in modified gravities with spatial covariance and of crucial importance in the construction for a power-counting renormalizable theory of quantum gravity \cite{horava1, horava6}.
 
 \subsubsection{Parity violating birefringences}
 
 The coefficients $\alpha_{\nu}$ ($\beta_{\nu}$) and $\alpha_{\mu}$ ($\beta_{\mu}$) represent the parity violating effects on the propagation of the GWs. Among these parameters, $\alpha_{\nu}$ ($\beta_{\nu}$) label the effects of amplitude birefringence, while $\alpha_{\mu}$ ($\beta_{\mu}$) is related to the velocity birefringence. In this paper we will consider the  velocity birefringence effects only. With the GW events detected by LIGO and Virgo, the GW constraints on the parity-violating birefringence for $\beta_{\mu} = 1$ \cite{yi-fan1, Wang:2021gqm} and $\beta_\mu =-1$ \cite{ZHUTAO} have been obtained by comparing the modified waveforms with GW data. With these constraints, one is able to infer the gravitational wave constraints on the corresponding parity-violating gravities. In this paper, we shall consider the case with $\beta_{\mu}=3$ and derive the corresponding constraints by using the signals of GW events detected by LIGO-Virgo. This case corresponds to the fifth-order spatial derivatives terms in modified gravities with spatial covariance, which can arise naturally from a specific spatial covariant gravity \cite{Gao:2019liu} and Ho\v{r}ava-Lifshitz gravity \cite{horava5}.

 \section{Specific theories with high-order spatial derivatives}
 \renewcommand{\theequation}{3.\arabic{equation}} \setcounter{equation}{0}
 
In this section we present two specific theories that contain the high-order spatial derivatives in the framework of spatial covariant gravity.   

\subsubsection{Ho\v{r}ava-Lifshitz gravity}

The Ho\v{r}ava-Lifshitz (HL) gravity is based on the view that Lorentz symmetry is a symmetry only at low energy, but may not exist at high energy \cite{horava1}. Because of this feature, one can establish a high-energy gravity without Lorentz symmetry in the UV. This gravity theory contains higher-order spatial derivatives, and also keep the time derivative operators to the second-order. In this paper, we are going to focus on an extension of the HL gravity by abandoning the projectability condition but imposing an extra local U(1) symmetry that was proposed in \cite{ Zhu:2011yu}, in which the gravitational sector has the same degree of freedom as that in GR, i.e., only spin-2 massless gravitons exist.

The action of the HL gravity consists kinetic and potential parts. The kinetic part only includes at most second time derivatives, which then is given by \cite{horava1, horava6}
\bqn
S_{\rm k} = \frac{M_{\rm Pl}^2}{2} \int dt d^3 x \sqrt{g} N (K_{ij}K^{ij} - \lambda K^2),
\eqn
where $K_{ij}$ is the extrinsic curvature of the constant time hypersurface defined by $K_{ij} = (\dot g_{ij}- \nabla_i N_j - \nabla_j N_i)/2 N$ with $N$ and $N_i$ being the lapse function and shift vector respectively, $K$ is the trace of $K_{ij}$. In construction of the potential part, it is of crucial importance to consider the spatial derivative up to the sixth order, in order to make the theory to be power-counting renormalizable. In different versions of the HL theory, the potential part has different forms, but in general it can be cast into the sum of different orders of spatial derivatives, i.e., 
\bqn
S_{\rm V} = \frac{M_{\rm Pl}^2}{2} \int dt d^3 x \sqrt{g} N \mathcal{L}_{\rm V},
\eqn
where 
\bqn
\mathcal{L}_{\rm V} = \sum_{n=2} \mathcal{L}^{(n)}(N, N^i, g_{ij}),
\eqn
with $\mathcal{L}^{(n)}$ denoting the part of Lagrangian that contains operators of the $n$th-order spatial derivative only. In the above, $n$ is an even number for parity-conserving terms and is odd for parity violating terms. 

By abandoning the Lorentz symmetry, the HL theory also provides a natural way to incorporate the parity violation terms into the theory. For our current purpose, we consider the third- and/or fifth-order spatial derivative operators to the potential term ${\cal{L}}$ of the total action in \cite{horava4, horava5, Zhu:2011yu},
\bqn \label{parity action}
\Delta {\cal{L}} &=& \frac{1}{M_{\rm PV}^3} \left(\alpha_0 K_{ij} R_{ij} +\alpha_2 \varepsilon^{ijk} R_{il} \Delta_j R^l_k \right) \nb\\
&& + \frac{\alpha_1 \omega_3(\Gamma)}{M_{\rm PV}}+``\dots",
\eqn
here $M_{\rm PV}$ is the energy scale above which the high-order derivative operators become important. The coupling constant $\alpha_0,\;\alpha_1,\;\alpha_2$ are dimensionless and arbitrary, and $\omega_3(\Gamma)$  the 3-dimensional gravitational CS term.  ``..." denotes the rest of the fifth-order operators given in Eq.(2.6) of \cite{Zhu:2011yu}. Since they have no contributions to tensor perturbations, in this paper we shall not write them out explicitly. 

The general formulas of the linearized tensor perturbations were given in \cite{horava5}, so in the rest of this subsection we give a very brief summary of  the main results obtained there.  Consider a flat FRW universe and assuming that matter fields have no contributions to tensor perturbations, the quadratic part of the total action can be cast in the form,
\bqn
S^{(2)}_{\text{ g}}&=& \frac{M_{\rm Pl}^2}{2} \int d\eta d^3 x \Bigg\{\frac{a^2}{4} (h_{ij}')^2-\frac{1}{4} a^2 (\partial_k h_{ij})^2 \nb\\
&&\;\;\;  -\frac{\hat{\gamma_3}}{4 M_{\rm LV}^2}(\partial^2h_{ij})^2-\frac{\hat{\gamma}_5}{4 M_{\rm LV}^4 a^2}(\partial^2 \partial_k h_{ij})^2\nonumber\\
&&\;\;\;-\frac{\alpha_1 a \epsilon^{ijk}}{2 M_{\rm PV}} (\partial_l h_{i}^m \partial_m \partial_j h_k^l-\partial_l h_{im} \partial^l \partial_j h^m_k) \nb\\
&&\;\;\; -\frac{\alpha_2 \epsilon^{ijk}}{4 M_{\rm PV}^3 a}\partial^2 h_{il} (\partial^2 h^l_k)_{,j} - \frac{3 \alpha_0 {\cal{H}}}{8 M_{\rm PV} a }(\partial_k h_{ij})^2\Bigg\},\nb\\
\eqn
where $\hat{\gamma}_3  \equiv  ({2M_{\rm LV}}/{M_{\rm Pl}})^2  {\gamma}_3$ and $\hat{\gamma}_5 \equiv({2 M_{\rm LV}}/{M_{\rm Pl}})^4 {\gamma}_5$, and $\gamma_3$ and $\gamma_5$ are the dimensionless coupling constants of the theory. To avoid fine-tuning, ${{\alpha}}_{n}$ and  $\hat{\gamma}_{n}$  are expected to be  of  the same order.
Then, the field equations for $h_{ij}$ read,
 \bqn
 \label{DFE}
 && h''_{ij}+2\mathcal{H}h'_{ij}- \alpha^2 \partial^2 h_{ij}+ \frac{\hat\gamma_3}{a^2 M_{\rm LV}^2}\partial^4 h_{ij}-\frac{\hat\gamma_5}{a^4 M_{\rm LV}^4}\partial^6 h_{ij} \nb\\
 && \;\; \;+ \epsilon_{i}^{\;\; lk}\left(\frac{2\alpha_1}{a M_{\rm PV}} + \frac{\alpha_2}{a^3 M_{\rm PV}^3}\partial^2 \right) \left(\partial^2h_{jk}\right)_{,l} = 0,\nb\\
 \label{eq7}
 \eqn
where $\alpha^2 \equiv 1+ {3\alpha_0{\cal{H}}}/{(2 M_{\rm PV} a)}$. In the late universe, $a\sim 1$, and $\mathcal{H}\ll M_{\rm PV}$, so we find $\alpha^2\rightarrow 1$. 

To study the evolution of $h_{ij}$, we expand it over spatial Fourier harmonics. For each circular polarization mode, the equation of motion of GW is given by
 \bqn
 h_{\rm A}'' +2 \mathcal{H} h_{\rm A}'+\omega_{\rm A}^2 h_{\rm A}=0,
 \eqn
 with
\bqn
 \omega^2_{\rm A}(k, \eta) &\equiv& \alpha^2 k^2 \Bigg[1+ \delta_1 \rho_{\rm A} \frac{\alpha k}{ M_{\rm PV} a}+\delta_2 \left(\frac{\alpha k}{M_{\rm LV} a}\right)^2 \nb\\
 &&- \delta_3 \rho_{\rm A} \left(\frac{\alpha k}{ M_{\rm PV} a}\right)^3 + \delta_4 \left(\frac{\alpha k}{M_{\rm LV} a}\right)^4 \Bigg],\nb\\
 \eqn
where $\delta_1\equiv {2 \alpha_1 }/{\alpha^3 }$, $\delta_2\equiv {\hat{\gamma_3} }/{\alpha^4 }$, $\delta_3\equiv {\alpha_2 }/{\alpha^5 }$, $\delta_4\equiv {\hat{\gamma_5} }/{\alpha^6 }$. In comparison with the formula in Eq.(\ref{eom_A}), we find the coefficients in HL gravity,
\bqn
\mathcal{H} \bar{\nu} &=& 0, \\
 \bar{\mu}&=&\delta_2 \left(\frac{k}{a M_{\rm LV}}\right)^2+\delta_4 \left(\frac{k}{a M_{\rm LV}}\right)^4 \nb\\
 &&+\frac{3\alpha_0{\cal{H}}}{(2 M_{\rm PV} a)}, \lb{barmu}\\
 \mathcal{H} \nu_{\rm A}&=&0, \\
 \mu_{\rm A} &=&\delta_1 \rho_{\rm A} \left(\frac{k}{a M_{\rm PV}}\right)- \delta_3 \rho_{\rm A} \left(\frac{k}{a M_{\rm PV}}\right)^3,\lb{muAbar}
 \eqn
where we have considered the relation $\alpha^2\rightarrow1$. 

For the Lorentz violating effects, labeled by $\bar{\mu}$ and $\bar{\nu}$, the observational constraint on the effect of the first term (c.f. $\delta_2$ term) in (\ref{barmu}) from GWTC-1 and GWTC-2 has been derived in \cite{testGR_GWTC1, testGR_GWTC2}. For parity violating effects, labeled by $\mu_A$, the observational constraint on the effect of the first term in (\ref{muAbar}) has been considered in \cite{sai_wang, Wang:2021gqm, yi-fan1}. In this paper, we will not consider $\delta_1$ and $\delta_2$ terms, and concentrate on the effects of $\delta_3$ and $\delta_4$ on GWs and their constriants from GW data.

 \subsubsection{Spatial covariant gravity}
 
We first start with the general action of the spatial covariant gravity \cite{Gao:2020yzr, Gao:2019liu},
\bqn\lb{action}
S = \int dt d^3 x N \sqrt{g} {\cal L}(N, g_{ij}, K_{ij}, R_{ij}, \nabla_i, \varepsilon_{ijk}),\nb\\
\eqn
where $R_{ij}$ is the intrinsic curvature tensor, $\nabla_i$ denotes the spatial covariant derivative with respect to $g_{ij}$, and $\varepsilon_{ijk}=\sqrt{g} \epsilon_{ijk}$ is the spatial Levi-Civita tensor with $\epsilon_{ijk}$ being the total antisymmetric tensor. The most important feature of the spatial covariant gravity is that it is only invariant under the 3-dimensional spatial diffeomorphism, which breaks the time diffeomorphism. Normally, the violation of the time diffeomorphism can lead to extra degree of freedom, in addition to the two tensorial degree of freedom in GR. Indeed, it has been verified that the spatial covariant gravity described by the action (\ref{action}) can propagate up to 3 dynamical degrees of freedom.

There are a lot approaches to construct the gravitational theories with spatial covariance. In this paper, for simplicity, we will not write out the action explicitly and only give a brief summary of GWs in the spatial covariant gravity. In general, the equation of motion of $h_A$ in a spatial covariant gravity can be written in the form \cite{Gao:2019liu},
\bqn
h_A '' + (2+\Gamma_A){\cal H}h_A' + \omega_A^2 h_A =0,
\eqn
where
\bqn
\Gamma_A = \frac{1}{\cal H} \frac{\partial_\tau {\cal G}^{(A)}(\tau, k)}{{\cal G}^{(A)}(\tau, k)}, \\
\frac{\omega_A^2}{k^2} = \frac{{\cal W}^{(A)}(\tau,k)}{{\cal G}^{(A)}(\tau, k)},
\eqn
where ${\cal G}^{(A)}$ and ${\cal W}^{(A)}$ are functions of time and wave number $k$. Depending specific terms in the spatial covariant gravity, ${\cal G}^{(A)}$ and ${\cal W}^{(A)}$ can have different forms. In above expressions, $\Gamma_A$ can lead to amplitude modifications, while $\omega_A^2$ modifies phases of GWs. In this paper, we only focus on the phase modification on GWs thus we will not consider the effect of $\Gamma_A$ and their GW constraints. For $\omega_A^2$, the general form can be expressed as \cite{Gao:2019liu}
\bqn
\frac{\omega_A^2}{k^2} = \frac{{\cal W}_0 + \rho_A {\cal W}_1 \tilde k + {\cal W}_2 \tilde k^2 + \rho_A {\cal W}_3 \tilde k^3 + {\cal W}_4 \tilde k^4 + \cdots }{{\cal G}_0 + \rho_A {\cal G}_1 \tilde k + {\cal G}_2 \tilde k^2 + \rho_A {\cal G}_3 \tilde k^3 + {\cal G}_4 \tilde k^4 + \cdots }, \nb\\
\eqn
where ${\cal G}_n$ and ${\cal W}_n$ are functions of time. In the above expression, ${\cal G}_n$ and ${\cal W}_n$ for even number $n$ denote the Lorentz violating effects, while for odd number are effects from parity violating operators. 

The observational constraints on the terms with coefficients ${\cal W}_1$, ${\cal W}_2$, and ${\cal G}_1$ by using GW data can be inferred directly from the analysis in \cite{testGR_GWTC1, testGR_GWTC2, sai_wang, tanaka, yi-fan1, Wang:2021gqm}. In this paper, we will focus on the observational constraints on ${\cal W}_3$ and ${\cal W}_4$, which can arise from the fifth-order and sixth-order spatial derivative terms in the action of gravity with spatial covariance. 

\section {Bayesian inference for GW data}
 \renewcommand{\theequation}{4.\arabic{equation}} \setcounter{equation}{0}
 
 \subsection{Bayesian inference for GW data}
 
 In this paper we perform  the Bayesian inference to  constraint the parity and Lorentz violations with selected GW events from the LIGO-Virgo catalogs GWTC-1 and GWTC-2. Bayesian analysis is also often used in GW  parameter estimation such as chirp mass, redshift, right ascension and luminosity distance etc. of GW source. When we have GW data $d$, we compare the GW data with the parameter $\vec {\theta}$ inferred by the instrument with the data of  LIGO-Virgo catalogs  to infer the distribution of the parameter $\vec{\theta} $. Next, we write the Bayesian theorem in the context of GW astronomy:
 \bqn
 P({\vec \theta}|d, H)=\frac{P(d| \vec{\theta}, H) P(\vec{\theta}| H)}{P(d|H)},\lb{bayes}
 \eqn
 where $H$ is the waveform model, $P(\vec{\theta}| H)$ is the prior distribution for model parameters $\vec{\theta}$ and $P(d| \vec{\theta}, H) $ is the likelihood  given a specific set of model parameters,  $P(d|H)$ is normalization factor called the ‘evidence’, 
 \bqn
 P(d|H) \equiv \int d \vec{\theta} P(d| \vec{\theta}, H) P(\vec{\theta}| H),
 \eqn
and  $ P({\vec \theta}|d, H)$ denotes the posterior probability distributions for physical parameters ${\vec{\theta}}$ which describe the observational data.

 In most cases, we use the matched filtering method to extract the signal from the noise (we assume that the noise is Gaussian white noise) because the GW signal from compact binary combination is very weak compared with most other signals, such as electromagnetic wave signals. The likelihood function of the  matched filtering method could  be defined in the following form:
 \bqn
 P(d|{\vec \theta}, H) \propto \prod_{i=1}^{n} e^{-\frac{1}{2}\langle d_i-h({\vec \theta})|d_i-h({\vec \theta})\rangle} ,
 \eqn
 where $h(\vec{\theta})$ is the GW waveform template for model $H$ and $i$ represents the $i$ th GW detector. The noise weighted inner product $\langle A|B \rangle$ is defined as
 \bqn
 \langle A|B \rangle = 4\; {\rm Re} \left[\int_0^\infty \frac{A(f) B(f)^*}{S(f)} df\right],
 \eqn
 where $\;^*$ denotes complex conjugation and $S(f)$ is the power spectral density (PSD) function of the detector.

Next, we  restrict  our  attention to the  cases  with  Lorentz and  parity violations due  to high-spatial derivative terms in gravitational actions. We utilize the Python package \texttt{BILBY} to perform the Bayesian inference by analyzing the GW data from selected 47 events of BBH and BNS mergers in the LIGO-Virgo catalogs GWTC-1 and GWTC-2. We use the waveform template given in (\ref{hplus}) and (\ref{htimes}) with $\delta \Psi_2$ for the Lorentz violating effect and $\delta \Psi_1$ for parity violating effect respectively. We employ template \texttt{IMRPhenomPv2} for the GR waveform $h^{\rm GR}_{+, \; \times}(f)$ for BBH events except the event GW190814 and GW190521, and \texttt{IMRPhenomPv2\_NRTidal} for BNS events. For GW190814 and GW190521, we employ the waveform \texttt{IMRPhenomXPHM} which includes the subdominant harmonic modes of GW and accounts for spin-precession effects for a quasicircular-orbit binary black hole coalescence. The explanation of parameter vector $\vec{\theta}$ for these waveform templates can be found in \cite{ZHUTAO,gwtc1}.

We have constructed modified GW waveform through Eq. (\ref{hplus}) and Eq. (\ref{htimes}) for $\beta_{ \mu} =3$ and $\beta_{\bar \mu} =4$ with respect to parity violation and Lorentz violation respectively. We ignore the contribution of $\delta h_1$ and $\delta h_2$ because their contributions to the GW strain are negligibly small compared to phase modifications from $\delta \Psi_1$ and $\delta \Psi_2$. For our purpose we consider the Lorentz violation and parity violation effects separately, which means when we consider the Lorentz violating effects in the waveform, we have set the contributions of parity violation to vanish, and vice-versa. When $\beta_{ \mu}$ =3 and $\beta_{\bar \mu}$ =4, the phase modification $\delta \Psi_1$ and $\delta \Psi_2$ are written respectively in the following forms,
\bqn
\delta \Psi_1 &=& A_\mu {(\pi f)}^4, \lb{deltaPsi}\\
\delta \Psi_2 &=& A_{\bar \mu} (\pi f)^5, \lb{deltaPsi2}
\eqn
where 
\bqn
A_\mu &=& \frac{2}{(M_{\rm PV})^3}  \int_{t_e}^{t_0}\frac{\alpha_\mu}{a^4}dt,
\eqn
and
\bqn
A_{\bar \mu} &=& \frac{16}{5 (M_{\rm LV})^4}  \int_{t_e}^{t_0}\frac{\alpha_{\bar \mu}}{a^5}dt,
\eqn
Then we consider a series of GW events composed of data $\{d_i\}$, described by parameters $\{{\vec \theta_i}\}$, then for each event, posterior for all the parameters describing the event can be written as
\bqn
P(A_\mu, {\vec \theta_i} | d_i) = \frac{P(M_{\rm PV}, \vec{\theta}_i) P(d_i| M_{\rm PV}, \vec{\theta}_i)}{P(d_i)},
\eqn
for testing the parity violating waveform, and
\bqn
P(A_{\bar \mu}, {\vec \theta_i} | d_i) = \frac{P(M_{\rm LV}, \vec{\theta}_i) P(d_i| M_{\rm LV}, \vec{\theta}_i)}{P(d_i)},
\eqn
for testing Lorentz violating waveforms.
 
\subsection{Selection of data samples}

Up to now,  the LIGO-Virgo catalogs GWTC-1 and GWTC-2 have found and reported a total of 50 gravitational wave events. 
However, it is mentioned in \cite{LIGOScientific:2021usb} that three of GWTC-2 events, the events GW190424\_180648, GW190426\_152155, and GW190909\_114149, have a probability of astrophysical origin $p_{\rm astro}<0.5$, so we carefully exclude them and then select 47 BBH and BNS events in our analysis as presented in Table.~\ref{1} (see also in Table.~\ref{2}.
The data of these 47 GW events are downloaded from Gravitational Wave Open Science Center \cite{data_GW}. 
For these events, we perform parameter estimation using Bayesian analysis by selecting $4 \;{\rm s}$ or $8 \; {\rm s}$ or $16 \; {\rm s}$  data for BBH and $128\; {\rm s}$ data for BNS over the selected GW parameters $\vec{\theta}$ and the parity violating parameter $A_\mu$ (the Lorentz violating parameter $A_{\bar \mu}$). The prior for the standard GW parameters ${\vec{\theta}}$ are consistent with those used in \cite{gwtc1, gwtc2}. The prior for $A_\mu$ and  $A_{\bar \mu}$  are chosen to be uniformly distributed. We use the package {\texttt BILBY}  \cite{bilby} to perform the Bayesian analysis and the posterior distribution is sampled by the nest sampling method dynesty over the fiducial BBH and BNS data and the parity violating parameter $A_\mu$ and the Lorentz violating parameter $A_{\bar \mu}$. We report our main results in the next section.

 \begin{table}
\caption{90\% credible level lower bounds on $M_{\rm PV}$ from the Bayesian inference by  analyzing the selected 46 GW events in the LIGO-Virgo catalogs GWTC-1 and GWTC-2. Note that  result from GW190521 is excluded.}
\label{1}
\begin{ruledtabular}
\begin{tabular} {c|cc}
catalogs & GW events &  Lower limit [$10^{-15}\; {\rm GeV}$]\\
\hline 
\multirow{3}*{GWTC-1}  
& GW150914 & 1.9 \\
& GW151012 & 3.3 \\
& GW151226 & 5.9 \\
& GW170104 & 2.0  \\
& GW170608 & 5.8  \\
& GW170729 & 1.6  \\
& GW170809 & 2.1  \\
& GW170814 & 2.8  \\
& GW170817 & 7.5  \\
& GW170818 & 2.2  \\
& GW170823 & 2.0  \\
\hline
\multirow{16}*{GWTC-2} 
& GW190408\_181802&  3.2 \\
& GW190412 &   3.1 \\  
& GW190413\_052954&  2.0\\
& GW190413\_134308 & 0.9\\
& GW190421\_213856 & 1.5 \\
& GW190425 & 5.1 \\
& GW190503\_185404 & 1.7 \\
& GW190512\_180714 & 3.7 \\
& GW190513\_205428 & 2.1 \\
& GW190514\_065416 & 1.4 \\
& GW190517\_055101 & 2.6 \\
& GW190519\_153544 & 1.4 \\
& GW190521\_074359 & 2.4 \\
& GW190527\_092055 & 2.1 \\
& GW190602\_175927 & 0.9 \\
& GW190620\_030421 & 1.1 \\
& GW190630\_185205 & 2.4 \\
& GW190701\_203306 & 1.2 \\
& GW190706\_222641 & 0.6 \\
& GW190707\_093326 & 6.3 \\
& GW190708\_232457 & 3.9 \\
& GW190719\_215514 & 0.6 \\
& GW190720\_000836 & 5.4 \\
& GW190727\_060333 & 2.3 \\
& GW190728\_064510 & 7.3 \\
& GW190731\_140936 & 1.4 \\
& GW190803\_022701 & 1.4 \\
& GW190814 & 4.5 \\
& GW190828\_063405 & 3.0 \\
& GW190828\_065509 & 3.6 \\
& GW190910\_112807 & 1.5 \\
& GW190915\_235702 & 2.1 \\
& GW190924\_021846 & 8.0 \\
& GW190929\_012149 & 0.8 \\
& GW190930\_133541 & 6.0 \\
\hline
combined  &  &   10.2
\end{tabular}
\end{ruledtabular}
\end{table}
 
\section{Results of constraints on Lorentz and parity violations}
 \renewcommand{\theequation}{5.\arabic{equation}} \setcounter{equation}{0}

For 46 GW events in all the 47 events we analyzed, we do not find any significant signatures of Lorentz and parity violation due to the fifth and sixth-order spatial derivatives. To illustrate our results, we present briefly the marginalized posterior distributions of $A_\mu$ (upper panel) and $A_{\bar \mu}$ (bottom panel) in Fig.~\ref{violin} for those GW events with chirp mass ${\cal M} \lesssim 25 M_{\odot}$. In this figure, the region in the posterior between the upper and lower bar denotes the 90\% credible interval for all these events. From Fig.\ref{violin}, it is also shown that the relatively low-mass BBH and BNS events, such as GW170817, GW190425, GW190728\_064510, GW190924\_021846 give tighter constraints on both $A_\mu$ and $A_{\bar \mu}$. This is exactly what we expected, as the low-mass event can lead to larger phase corrections, as shown in (\ref{deltaPsi}) and (\ref{deltaPsi2}). For all the 46 events among the 47 events, our results show that the GR values $A_\mu = 0$ and $A_{\bar \mu}=0$ is well within the 90\% confidence level. 

The only one exception is the GW events GW190521. In Fig.~\ref{pdf_3}, we present the posterior distributions of $A_\mu$ and $A_{\bar \mu}$ for GW190521, which seems favor nonzero values for $A_\mu$ and $A_{\bar \mu}$. Similar results for constraining parity violation with event GW190521 has also been reported in \cite{Wang:2021gqm}. It is mentioned in \cite{Wang:2021gqm} that such result may be caused by the limitations of the existing waveform approximants, such as systematic errors during merger phase of the waveform, or by the existence of physical effects such as eccentricity which are not taken into account by the current waveform approximants. For this reason, we exclude this event in our analysis in the rest of this paper.

From the posterior distributions of $A_{\mu}$($A_{\bar \mu}$) and redshift $z$ calculated from the selected 46 events, one can convert $A_\mu$($A_{\bar \mu}$) and $z$ into $M_{\rm PV}$ ($M_{\rm LV}$). In Fig. \ref{pdf_45} we separately plot the marginalized posterior distribution of $M_{\rm PV}^{-3}$ and $M_{\rm LV}^{-4}$. Then the lower bounds on $M_{\rm PV}$ and $M_{\rm LV}$ for each individual event can be calculated from the corresponding posterior distributions. In Table-\ref{1}, we present the 90\% credible level lower bounds on $M_{\rm PV}$ from the Bayesian inference of individual event and in Table-\ref{2}, we present the 90\% credible level lower bounds on $M_{\rm LV}$ from the Bayesian inference of individual event. 

\begin{figure*} 
{
\includegraphics[width=16.1cm]{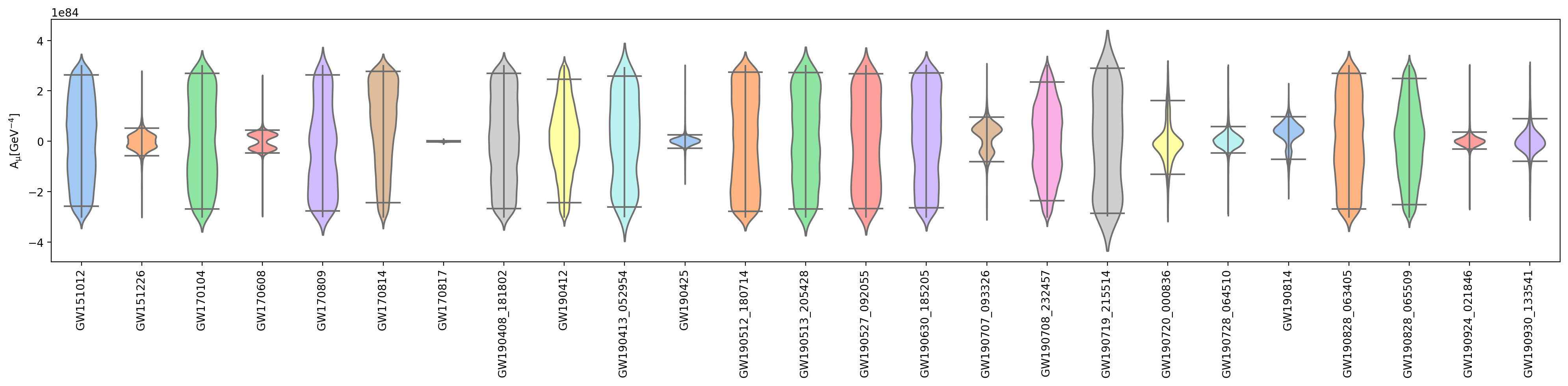}
\includegraphics[width=16.1cm]{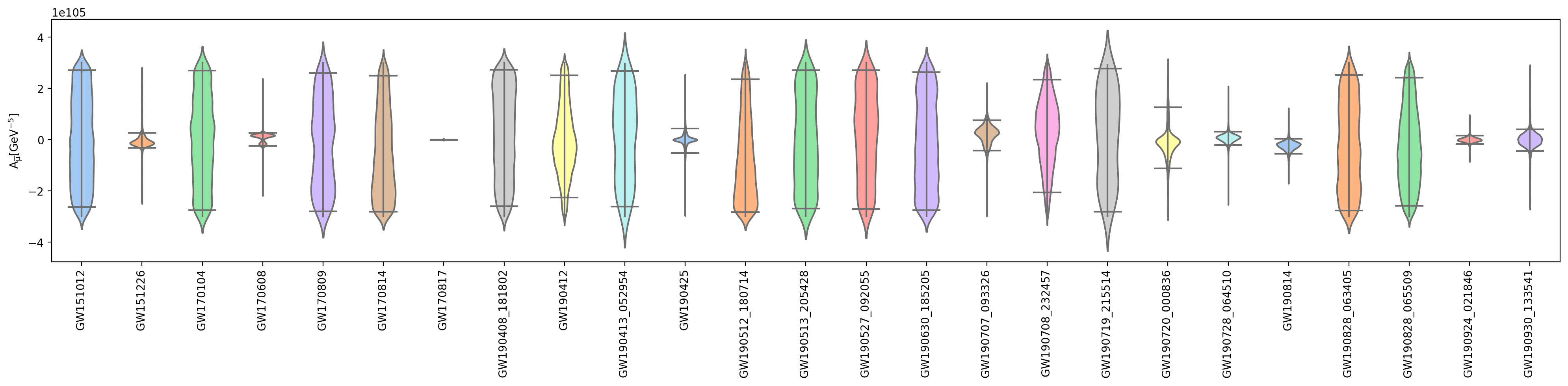}}
\caption{Violin plots of the posteriors of the parameter $A_\mu$ and $A_{\bar \mu}$. The results are obtained by analyzing the selected 25 GW events. The region in the posterior between the upper and lower bar denotes the $90\%$ credible interval. } 
\lb{violin}
\end{figure*}

\begin{figure*} 
{
\includegraphics[width=8.1cm]{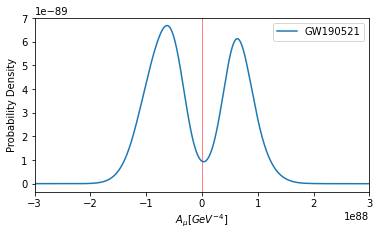}
\includegraphics[width=8.1cm]{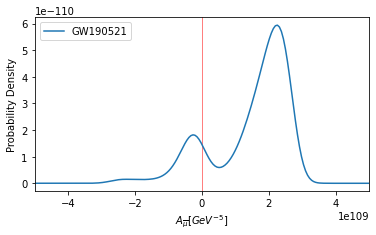}}
\caption{The distribution of posteriors of the parameter $A_\mu$ and $A_{\bar \mu}$ for GW150921.}
\lb{pdf_3}
\end{figure*}

The parameter $M_{\rm PV}$($M_{\rm LV}$) are universal quantities for all GW events. One can combine all the individual posterior of $M_{\rm PV}$ ($M_{\rm LV}$) for each event to get the overall constraint. This can be done by multiplying the posterior distributions of all these events together through 
\bqn
P(M_{\rm PV}|\{d_i\}, H) \propto \prod_{i=1}^{N} P(M_{\rm PV}| d_i, H),\\
P(M_{\rm LV}|\{d_i\}, H) \propto \prod_{i=1}^{N} P(M_{\rm LV}| d_i, H),
\eqn
where $d_i$ denotes data of the $i$ th GW event and N denotes selected quantities of data. We find that $M_{\rm PV}$ and $M_{\rm LV}$ can be constrained to be
\bqn
M_{\rm PV} > 1.0 \times 10^{-14} \; {\rm GeV},
\eqn
and
\bqn
M_{\rm LV} > 2.4 \times 10^{-16} \; {\rm GeV},
\eqn
at 90\% confidential level.

\section{conclusion and outlook}
 \renewcommand{\theequation}{5.\arabic{equation}} \setcounter{equation}{0}

 With the discovery of GWs from the coalescence of compact binary systems by LIGO/Virgo Collaboration, the testing of gravity in the strong gravitational fields becomes possible. In this paper, we consider the tests of Lorentz and parity violations arising from high-order spatial derivatives of gravity with spatial covariance by using the selected GW events in the LIGO-Virgo catalogs GWTC-1 and GWTC-2. For this purpose we calculate in detail the effects of the modified dispersion relation and velocity birefringence due to the Lorentz and parity violations on the GW waveforms. Decomposing the GWs into the left-hand and right-hand circular polarization modes, we find that the GW effects of both the Lorentz and parity violation can be explicitly presented by the modifications in the GW amplitude and phase.  We also mapped such amplitude  and  phase  modification  to  the  parametrized description  of  Lorentz  and parity  violating  waveforms proposed in \cite{waveform}.  
 
 With the modified waveforms, we perform  the  full  Bayesian  inference  with  the  help  of  the open source software \texttt{BILBY} on the selected 47 GW events in the LIGO-Virgo catalogs GWTC-1 and GWTC-2.  From our analysis, we do not find any signatures of Lorentz violation due to the sixth-order derivatives and parity violation due to the fifth-order spatial derivatives in the gravitational action with spatial covariance.  We thus place lower bounds on the energy scale $M_{\rm LV} > 2.4 \times 10^{-16} \; {\rm GeV}$ for Lorentz violation and $M_{\rm PV} > 1.0 \times 10^{-14} \; {\rm GeV}$ for parity violation at 90\% confidence level. Both constraints represent the first constraints  on  the  fifth-  and  sixth-order  spatial derivative terms respectively in the spatial covariant framework by using the observational data of GWs. Note that in deriving the above conclusions, we have excluded the result of GW190521 since its result may be significantly biased by waveform templates used in the analysis \cite{Wang:2021gqm, testGR_GWTC2}.
 
 We observe that the relatively low-mass BBH and BNS events, such as GW170817, GW190425, GW190728\_064510, GW190924\_021846 give tighter constraints. Since the next generation of GW detectors can detect lighter and more distant BBH and BNS events, it is expected such system can lead to more tighter constraints on $M_{\rm PV}$ and $M_{\rm LV}$ in the future. 

\section*{Acknowledgments}
T.Z., Q.W., and A.W. are supported in part by the National Key Research and Development Program of China Grant No. 2020YFC2201503, and the Zhejiang Provincial Natural Science Foundation of China under Grant No. LR21A050001 and No. LY20A050002, the National Natural Science Foundation of China under Grant No. 11675143 and No. 11975203, and the Fundamental Research Funds for the Provincial Universities of Zhejiang in China under Grant No. RF-A2019015.
R.N. and W.Z. are supported by NSFC Grants No. 11773028, No. 11633001, No. 11653002, No. 11421303, No. 11903030, the Fundamental Research Funds for the Central Universities, and the Strategic Priority Research Program of the Chinese Academy of Sciences Grant No. XDB23010200. J.C is supported by the Fundamental Research Funds for the Central Universities of China (No. N180503018). X.Z. is supported by the National Natural Science Foundation of China (Grants No. 11975072, No. 11835009, and No. 11690021), the Liaoning Revitalization Talents Program (Grant No. XLYC1905011), and the National 111 Project of China (Grant No. B16009).

\begin{table}
\caption{90\% credible level lower bounds on $M_{\rm LV}$ from the Bayesian inference by  analyzing the selected 46 GW events in the LIGO-Virgo catalogs GWTC-1 and GWTC-2. Note that result from GW190521 is excluded.}
\lb{2}
\begin{ruledtabular}
\begin{tabular} {c|cc}
catalogs & GW events &  Lower limits [$10^{-17}\; {\rm GeV}$]\\
\hline 
\multirow{3}*{GWTC-1} 
& GW150914 & 4.9 \\
& GW151012 & 8.1 \\
& GW151226 & 15.7 \\
& GW170104 & 5.1  \\
& GW170608 & 15.6  \\
& GW170729 & 3.9  \\
& GW170809 & 5.1  \\
& GW170814 & 6.9  \\
& GW170817 & 21.1  \\
& GW170818 &5.2 \\
& GW170823 & 4.8  \\
\hline
\multirow{16}*{GWTC-2} 
& GW190408\_181802&  7.6 \\
& GW190412 &   8.5 \\  
& GW190413\_052954&  4.4\\
& GW190413\_134308 & 2.3\\
& GW190421\_213856 & 3.5 \\
& GW190425 & 11.1 \\
& GW190503\_185404 & 3.9 \\
& GW190512\_180714 & 9.0 \\
& GW190513\_205428 & 5.2 \\
& GW190514\_065416 & 3.8 \\
& GW190517\_055101 & 6.5 \\
& GW190519\_153544 & 3.5 \\
& GW190521\_074359 & 5.6 \\
& GW190527\_092055 & 5.3 \\
& GW190602\_175927 & 2.1 \\
& GW190620\_030421 & 2.6 \\
& GW190630\_185205 & 6.8 \\
& GW190701\_203306 & 2.9 \\
& GW190706\_222641 & 2.9 \\
& GW190707\_093326 & 15.6 \\
& GW190708\_232457 & 9.8 \\
& GW190719\_215514 & 2.2 \\
& GW190720\_000836 & 13.3 \\
& GW190727\_060333 & 5.2 \\
& GW190728\_064510 & 19.7 \\
& GW190731\_140936 & 3.5 \\
& GW190803\_022701 & 3.4 \\
& GW190814 & 12.9 \\
& GW190828\_063405 & 7.1 \\
& GW190828\_065509 & 9.7 \\
& GW190910\_112807 & 3.8 \\
& GW190915\_235702 & 6.1 \\
& GW190924\_021846 & 20.5 \\
& GW190929\_012149 & 1.8 \\
& GW190930\_133541 & 17.3 \\
\hline
combined  &  &   24.0
\end{tabular}
\end{ruledtabular}
\end{table}

\begin{figure*} 
{
 \centering
\includegraphics[width=8.1cm]{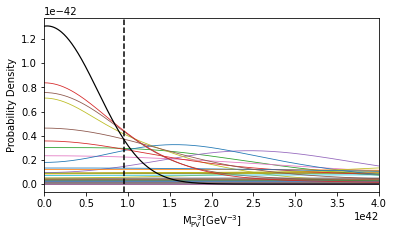} 
\includegraphics[width=8.1cm]{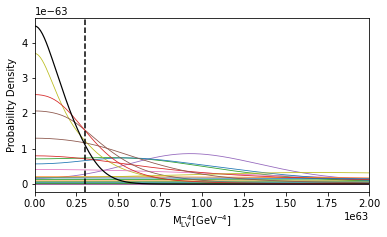}}\\
\caption{The posterior distributions for $M_{\rm PV}$ and $M_{\rm LV}$ from selected 46 GW events in the LIGO-Virgo catalogs GWTC-1 and GWTC-2.  The vertical dash line denotes the 90\% upper limits for $M_{\rm PV}$ ($M_{\rm LV}$) from combined results. Note that result from GW190521 is excluded.} 
\lb{pdf_45}
\end{figure*}

\appendix

\end{document}